# Harmonics of femtosecond and continuous-wave radiation for precision spectroscopy: generation efficiency and spectral line shape

N.O. Zhadnov[1, *], O.V. Khronusova[1,2], A.M. Russkikh[1,2], N.N. Kolachevsky[1,3], A.V. Masalov[1,3]

[1] *P.N. Lebedev Physical Institute of the Russian Academy of Sciences, 53 Leninskiy Prospekt, Moscow 119991, Russia*
[2] *Moscow Institute of Physics and Technology, 9 Institutskiy Pereulok, Dolgoprudny, Moscow Region 141701, Russia*
[3] *Russian Quantum Center, 30 Bolshoy Boulevard, bld. 1, Skolkovo Innovation Center Territory, Moscow 121205, Russia*
* *nik.zhadnov@yandex.ru*

*Precision spectroscopy in the UV and VUV ranges, including spectroscopy of the thorium-229 nuclear transition (148.4 nm), requires narrow-band sources obtained by frequency multiplication of stabilized lasers. Upon harmonic generation, the radiation intensity drops and the phase noise grows. In this work, both factors are analyzed for two types of primary sources: a continuous-wave single-frequency laser and a femtosecond optical frequency comb. It is shown that in the weak-conversion regime for equal average input powers, the power of the central mode in the spectrum of the Kth harmonic of femtosecond radiation matches the harmonic power of continuous-wave radiation for K = 2 and noticeably exceeds it for K > 2. Within the first-order dispersion approximation, group-velocity mismatch reduces the total conversion efficiency without affecting the power of the exactly phase-matched central mode. The transformation of the phase noise of an ultrastable laser upon harmonic generation leads to two effects that depend on K: the harmonic linewidth grows linearly, whereas the power fraction in the carrier decreases as a Gaussian function and is determined by the high-frequency noise of the stabilization loop (the servo bumps). Measurements of the phase-noise spectra of a laser stabilized to a Fabry–Pérot cavity show that for a typical rms phase excursion of ~100 mrad the carrier loses about half of its power already at the eighth harmonic, while for a non-optimal stabilization-loop gain carrier collapse occurs at the second–third harmonic.*

## **I.** Introduction

The vacuum ultraviolet (VUV) and deep-UV ranges contain a number of transitions of fundamental importance for physics and metrology: the Lyman series of atomic hydrogen [1,2], transitions in noble gases [3,4], cooling and clock transitions of the $Hg^+$ (194 and 282 nm) [5] and $Al^+$ (167 and 267 nm) [6] ions, as well as the only nuclear transition accessible to laser excitation, in the $^{229}$Th isotope with a wavelength of $\lambda$ = 148.4 nm [7–10], regarded as a frequency reference for a nuclear optical clock [11].

Precision spectroscopy places demanding requirements on the linewidth and power of laser radiation. Excitation of spectrally narrow clock transitions usually requires continuous-wave sources with a linewidth of the order of 1 Hz and a power of about a microwatt [6] or more. Pulsed VUV sources - free-electron lasers [12] and excimer lasers [13] - do not meet these requirements because of their broad spectral line, so the only remaining route is nonlinear-optical frequency conversion of stabilized lasers.

The principal difficulty in accessing the VUV range is that it offers virtually no media for direct continuous-wave laser generation, while the choice of transparent nonlinear materials is extremely limited. For harmonic generation in the visible and near-UV ranges (down to ~200 nm) there exists a well-developed set of crystals - LBO, BBO, KDP, CLBO - providing high conversion efficiency with phase matching based on birefringence. Below 200 nm, the birefringence of most crystals can no longer compensate the dispersion, which grows near the absorption edge, and the number of suitable materials shrinks sharply: birefringent phase matching for harmonic generation has been demonstrated in KBBF [14] and $NH_4B_4O_6F$ [15], but these crystals are opaque at 148.4 nm. Further advance into the VUV runs into a conflict between transparency and nonlinearity: transparency requires a large band gap, and as the band gap grows the nonlinear susceptibility falls rapidly. For crystals transparent at 148.4 nm, birefringent phase matching is unattainable, and quasi-phase-matching is required - a periodic or random modulation of the sign of the nonlinearity. Along this route, a continuous-wave power of about 1 nW has been obtained in an $SrB_4O_7$ crystal with a random domain structure [16], and recently, in a periodically poled crystal, a conversion efficiency of $7\cdot10^{-4}$ and a power of 40 μW have been achieved for comb conversion (1 nW per mode), corresponding to an estimate of ~60 nW in continuous-wave operation [17].

An alternative to solid-state media is provided by gaseous media, which are free from transparency and damage-threshold limitations. Their inversion symmetry forbids even-order processes, so only odd harmonics can be generated in them: the seventh harmonic of femtosecond radiation in xenon [18] or four-wave mixing in metal vapors, which has yielded the highest continuous-wave power at 148.4 nm to date of more than 100 nW [19]. The price is extremely low conversion efficiency, requiring high peak intensities or resonant enhancement, and a highly complex setup. In all the approaches listed, the efficiency of the final conversion into the VUV does not exceed $10^{-3}$-$10^{-4}$ and falls rapidly with increasing harmonic order, which justifies the undepleted-pump approximation used below.

Besides the power losses, K-fold frequency multiplication - regardless of its efficiency - multiplies the phase of the original radiation by K, and the phase-noise power spectral density by $K^2$ [20]. Thus, the route into the VUV via high harmonics poses two independent problems: the attainable power and the spectral purity. The primary source for subsequent conversion can be either a continuous-wave laser or a femtosecond optical frequency comb [21]; both types are stabilized to external references down to subhertz linewidths [22]. Since the mode structure of the comb is preserved upon harmonic generation [20], an individual mode of the converted spectrum can be used for spectroscopy. At first sight, the comb clearly loses to the continuous-wave laser: at equal average power, each of its modes carries $N \sim 10^5$ times less power, as noted in [23]. However, the high peak power of the pulses greatly enhances the efficiency of nonlinear conversion, and the question of which source provides more power in a single spectral component of the Kth harmonic requires quantitative analysis.

This work considers two aspects of frequency multiplication that determine the suitability of a source for precision spectroscopy: the power in the useful spectral component and the preservation of spectral purity. Section II compares the power of an individual mode of the Kth harmonic for continuous-wave and femtosecond radiation, considering the finite phase-matching bandwidth of the nonlinear medium. Section III analyzes the transformation of the spectrum of an ultrastable laser upon harmonic generation: line broadening and suppression of the carrier power fraction relative to the high-frequency regions of the spectrum.

**II.** Power of the modes of the converted radiation

In this section, we estimate the power of individual modes of the radiation produced by conversion into the $K$th harmonic, both for continuous-wave radiation and for a train of femtosecond pulses. We assume that both types of radiation are converted in the same nonlinear medium, with exact phase matching at the central frequency. The weak-conversion regime is considered, in which the pump is not depleted. Under these conditions, the quantity to be compared is the power of a single spectral component of the Kth harmonic: for a continuous-wave laser this is the entire harmonic power, while for the comb it is the power of the central mode of the converted spectrum, suitable for spectroscopy of a narrow transition.

In analyzing the conversion of radiation into the $K$th harmonic, we use the standard equation of nonlinear optics for the slowly varying amplitude of the $K$th-harmonic field $A(z,t)$ with central frequency $K\omega_0$:

$$\left(\frac{\partial}{\partial z}A + \frac{1}{v}\frac{\partial}{\partial t}A\right)e^{-iK\omega_0 t + ikz} = i\frac{2\pi\omega}{cn}\chi^{(K)}\left(\sum_m A_m e^{-i\omega_m t + ik_m z}\right)^K \quad (1)$$

where $k$ and $v$ are the wavevector of the harmonic field and the group velocity of the harmonic radiation in the medium: $v = \omega/k$, and $k_m$ and $v_m$ are the same for the modes of the input field[a]. The input radiation is represented here as a sum of mode fields $E(t) = \sum_m A_m e^{-i\omega_m t}$ with equidistant frequencies $\omega_m = \omega_0 + m\Omega$, where $m = \ldots, -2, -1, 0, 1, 2, \ldots$ and $\Omega$ is the mode spacing. Equation (1) is suitable for describing the conversion of both continuous-wave radiation and femtosecond-pulse radiation. In the case of continuous-wave radiation, only one mode has a nonzero amplitude $E(t) = A_0 e^{-i\omega_0 t}$ with radiation power $P_0 = A_0^2$. In the case of a femtosecond pulse train, the amplitudes of many modes share a common phase and a bell-shaped spectral distribution. In this case we will use the following expression for the mode amplitudes: $A_m = A_0\, exp\left(-2(\frac{\omega_m - \omega_0}{\Delta\omega})^2\right)$, where $\Delta\omega$ is the full width of the Gaussian power spectrum at the $1/e$ level. For an equidistant set of modes, the field of a single input pulse has the form

$$E(t) \approx A_0 e^{-i\omega_0 t}\int exp\left(-2(\frac{\omega-\omega_0}{\Delta\omega})^2\right)e^{-i(\omega-\omega_0)t}\frac{d\omega}{\Omega} = A_0 e^{-i\omega_0 t}\sqrt{\frac{\pi}{2}}\frac{\Delta\omega}{\Omega}\,exp\left(-\frac{(t\Delta\omega)^2}{8}\right), \quad (2)$$

where the factor $1/\Omega$ corresponds to replacing the sum with an integral. The pulse energy $\int |E(t)|^2 dt = A_0^2\pi\sqrt{\pi}\frac{\Delta\omega}{\Omega^2}$, divided by the repetition period $T = 2\pi/\Omega$, sets the radiation power

$$P_M = A_0^2\frac{\sqrt{\pi}}{2}\frac{\Delta\omega}{\Omega}. \quad (3)$$

The power of the central mode in the spectrum of the femtosecond radiation, $p_{max} = A_0^2 = \frac{2}{\sqrt{\pi}}\frac{\Omega}{\Delta\omega}P_M$, constitutes a small fraction of the continuous-wave power at equal total powers.

Equation (1) is written in the slowly-varying-amplitude approximation. On the left-hand side, the frequency dependence of the harmonic wavevector is assumed to be limited to the first-order approximation:

[a] The result of the derivation is Eq. (15); a reader interested only in the outcome may proceed directly to it.

$$k(\omega) \approx k(K\omega_0) + \left(\frac{dk}{d\omega}\right)_{\omega=K\omega_0} (\omega - K\omega_0) = k(K\omega_0) + \frac{1}{v}(\omega - K\omega_0), \quad (4)$$

where the second-order term $\frac{1}{2}\left(\frac{d^2k}{d\omega^2}\right)_{\omega=K\omega_0} (\omega - K\omega_0)^2$ and higher-order terms are neglected. In this case, the transition in Eq. (1) to the so-called retarded time $(t - z/v) \to t$ is justified, and the equation takes a form without the time derivative

$$\frac{d}{dz}A(t,z) = i\frac{2\pi\omega}{cn}\chi^{(K)}e^{i\omega t - ikz + iK\omega_0 z/v}\left(\sum_m A_m e^{-i\omega_m(t+z/v)+ik_m z}\right)^K, \quad (5)$$

where $k = k(K\omega_0)$. In the subsequent calculation we use a similar approximation for the wavevectors of the modes of the input radiation:

$$k_m \approx k(\omega_0) + \left(\frac{dk}{d\omega}\right)_{\omega=\omega_0} (\omega_m - \omega_0) = k_0 + \frac{1}{v_0}m\Omega. \quad (6)$$

Then

$$\frac{d}{dz}A(t,z) = i\frac{2\pi K\omega_0}{cn}\chi^{(K)}\left(\sum_m A_m e^{-im\Omega t + im\Omega z(1/v_0 - 1/v)}\right)^K \quad (7)$$

Here the phase-matching condition is assumed to hold for the central frequency of the spectrum: $k = Kk_0$. In the estimates below we assume a low conversion efficiency, at which the input radiation is not attenuated, so the input field amplitudes $A_m$ do not depend on the propagation coordinate *z*.

**A. *Continuous-wave laser.*** In this case, a single mode is converted into the harmonic, $A_0$:

$$\frac{d}{dz}A(z) = i\frac{2\pi K\omega_0}{cn}\chi^{(K)}A_0^K, \quad (8)$$

and at the output of a medium of length *L* we obtain the expression for the amplitude of the harmonic mode

$$A(L) = i\frac{2\pi K\omega_0}{cn}\chi^{(K)}A_0^K L. \quad (9)$$

The square of this amplitude gives the harmonic radiation power

$$p_0^{(K)} = \left(\frac{2\pi K\omega_0}{cn}\chi^{(K)}L\right)^2 P_0^K \quad (10)$$

This quantity is to be compared with the powers of individual modes in the harmonic spectrum produced from femtosecond radiation.

**B. *Femtosecond optical frequency comb.*** The sum in parentheses in Eq. (7) is a train of harmonic pulses propagating in the medium with group velocity *v*, with the pulse shape given by the *K*th power of the shape of the input pulses:

$$\left(\sum_m A_m e^{-im\Omega t + im\Omega z(1/v_0 - 1/v)}\right)^K = E^K\left(t - z\left(\frac{1}{v_0} - \frac{1}{v}\right)\right). \quad (11)$$

Considering (2), this sum in the spectral representation has the form

$$E^K\left(t - \left(\frac{z}{v_0} - \frac{z}{v}\right)\right) = \left(\sqrt{\frac{\pi}{2}}\frac{\Delta\omega}{\Omega}\right)^{K-1} A_0^K\sqrt{\frac{1}{K}}\sum_m e^{-im\Omega(t - z(1/v_0 - 1/v))}\, exp\left(-2\left(\frac{m\Omega}{\Delta\omega}\right)^2/K\right). \quad (12)$$

Integrating Eq. (7) with the field (12), we obtain the harmonic field at the output of the medium:

$$A(t,L) = i\frac{2\pi K\omega_0}{cn}\chi^{(K)}L\left(\sqrt{\frac{\pi}{2}}\frac{\Delta\omega}{\Omega}\right)^{K-1} A_0^K\sqrt{\frac{1}{K}}\times$$

$$\times\sum_m e^{-i\omega_m t}\, sinc\left(\frac{m\Omega L}{2}\left(\frac{1}{v_0} - \frac{1}{v}\right)\right) exp\left(-2\left(\frac{m\Omega}{\Delta\omega}\right)^2/K\right), \quad (13)$$

where the standard function $sinc\, x = \frac{\sin x}{x}$ is used. The spectrum of the field (13) is formed as a product of two factors: the idealized field $\propto exp\left(-2\left(\frac{m\Omega}{\Delta\omega}\right)^2/K\right)$, which is free of the phase-matching-bandwidth restriction, and the factor $sinc\left(\frac{m\Omega L}{2}(\frac{1}{v_0}-\frac{1}{v})\right)$, which accounts for the finite phase-matching bandwidth (Fig. 1). It is seen that the modes of the harmonic spectrum near exact phase matching ($m = 0$) are unaffected by the first-order dispersion of the medium, $\left(\frac{dk}{d\omega}\right)_{\omega=\omega_0}$ and $\left(\frac{dk}{d\omega}\right)_{\omega=K\omega_0}$, even though the overall conversion efficiency is reduced. The power of the central mode of the harmonic radiation is estimated as

$$p_{max}^{(K)} = \left(\frac{2\pi K\omega_0}{cn}\chi^{(K)}L\right)^2 \cdot \left(\sqrt{\frac{\pi}{2}}\frac{\Delta\omega}{\Omega}\right)^{2K-2} A_0^{2K} \cdot \frac{1}{K} =$$
$$=\left(\frac{2\pi K\omega_0}{cn}\chi^{(K)}L\right)^2 \cdot \left(\sqrt{\pi}\frac{\Delta\omega}{\Omega}\right)^{K-2} \cdot \frac{2}{K}P_M^K. \quad (14)$$

The ratio of the converted powers - femtosecond to continuous-wave - is (at equal total input powers $P_0 = P_M$):

$$\frac{p_{max}^{(K)}}{p_0^{(K)}} = \frac{2}{K}\left(\frac{\sqrt{\pi}\Delta\omega}{\Omega}\right)^{K-2} = \frac{2}{K}\left(\frac{2}{\sqrt{\pi}}\frac{T}{\tau}\right)^{K-2}, \quad (15)$$

where $\tau$ is the full pulse duration at the $1/e$ intensity level.

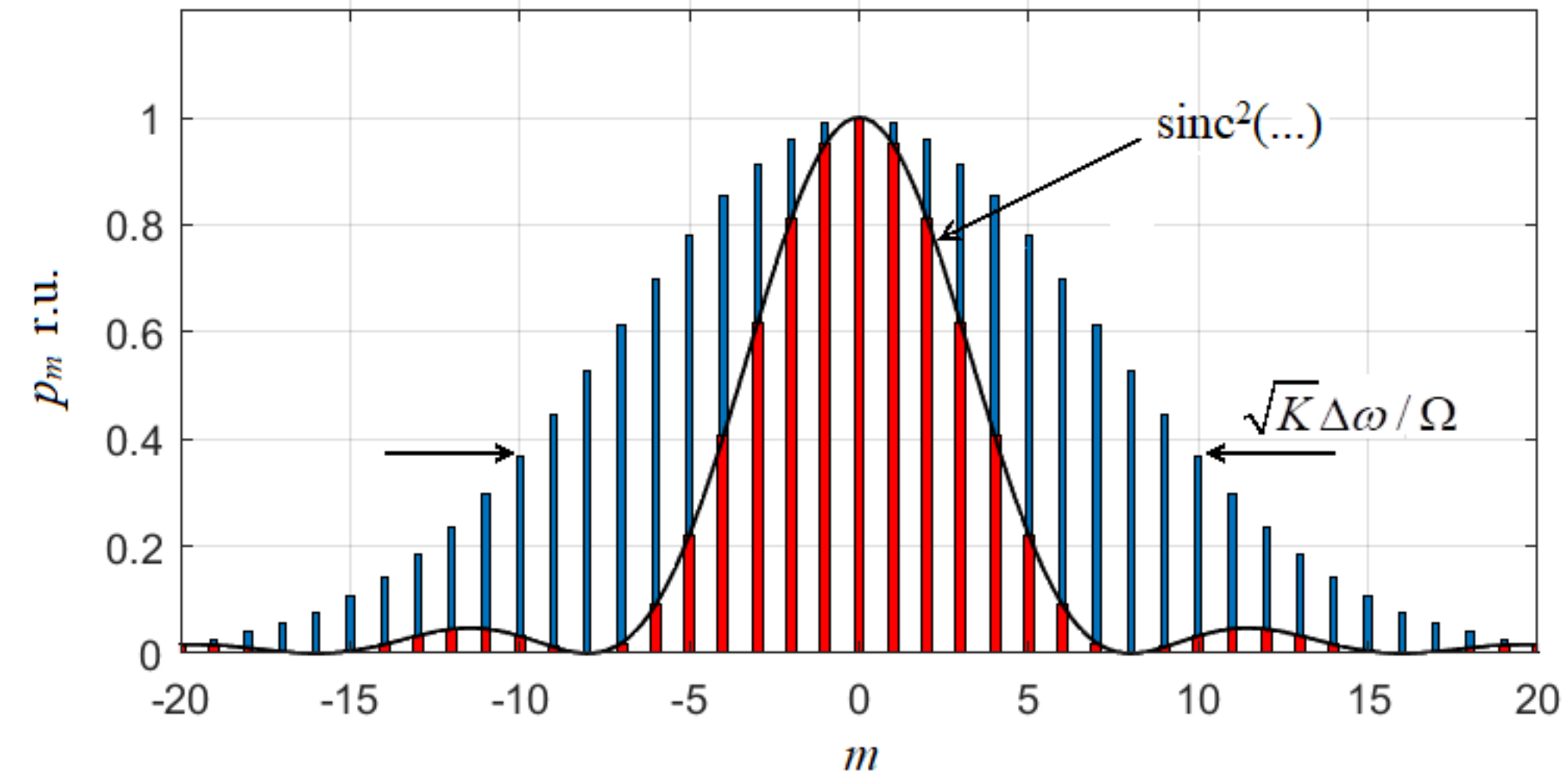

*Figure 1. Spectrum of the modes of the idealized harmonic radiation (12) (blue bars) and the profile of the sinc$^2$(...) function (black curve), which narrows the spectrum of the output modes (13) (red bars).*

In the case of second-harmonic generation ($K = 2$) the power of the central mode of the converted femtosecond radiation equals the harmonic power of the continuous-wave radiation $p_0^{(2)}$. For the third harmonic, the central mode of the femtosecond radiation noticeably exceeds the harmonic power of the continuous-wave radiation: $\frac{p_{max}^{(3)}}{p_0^{(3)}} = \frac{2\sqrt{\pi}}{3}\frac{\Delta\omega}{\Omega} = \frac{4}{3\sqrt{\pi}}\frac{T}{\tau}$. With each unit increase of the harmonic order, the ratio (15) grows by a factor of approximately $N = \Delta\omega/\Omega$ (the number of modes). A 100-femtosecond laser pulse with a period of $T \approx 10$ ns generates about $10^5$ modes, which can provide a substantial power gain in the generation of the third and higher harmonics.

The linear approximation of the medium dispersion may prove insufficient for correct estimates of the violation of the phase-matching conditions when the spectrum of the femtosecond radiation is broad. Let us estimate the range of validity of the linear approximation using the KBBF crystal as an example. Table 1 presents estimates of the contributions of the quadratic dispersion terms $d^2k/d\omega^2$ to the phase mismatch for the process of fourth-harmonic generation $\lambda = 742$ nm → $\lambda = 185.5$ nm for 100 fs and 1 ps pulses. The values of the second derivatives $d^2k/d\omega^2$ were calculated from the data of [24]. For each pulse duration, the table lists the spectral widths and the phase corrections to the spectral amplitudes of the input and converted radiation at the edges of the spectrum, $\frac{1}{2}\frac{d^2k}{d\omega^2}\left(\frac{\Delta\omega}{2}\right)^2$. For a crystal length of 1 mm, the phase deviations do not exceed 0.1 rad in all cases. For a crystal length of 1 cm and 100 fs pulses, these estimates indicate that second-order dispersion can no longer be neglected and that the first-order treatment requires refinement.

| | $\lambda$ | 742 nm | 185.5 nm |
|---|---|---|---|
| | $\frac{d^2k}{d\omega^2}$, fs$^2$/mm | 38.2 | 124.2 |
| $\tau = 1$ ps | $\Delta\omega$, fs$^{-1}$ | 0.004 | 0.008 |
| | $\frac{1}{2}\frac{d^2k}{d\omega^2}\left(\frac{\Delta\omega}{2}\right)^2$, rad/mm | $0.076\cdot10^{-3}$ | 0.001 |
| $\tau = 100$ fs | $\Delta\omega$, fs$^{-1}$ | 0.04 | 0.08 |
| | $\frac{1}{2}\frac{d^2k}{d\omega^2}\left(\frac{\Delta\omega}{2}\right)^2$, rad/mm | $7.6\cdot10^{-3}$ | 0.10 |

*Table 1 Estimates of second-order dispersive phase corrections at the spectral edges of the fundamental and fourth-harmonic fields in KBBF.*

**III.** Transformation of the ultrastable-laser line upon harmonic generation

In the process of nonlinear generation of the $K$th harmonic, the phase noise of the radiation changes. In the weak-conversion limit, the phase noise increases by a factor of K in accordance with the power-law nature of the conversion. This leads to a change in the shape of the spectrum and an increase in its width. In addition, the power fraction of the central component (the carrier) decreases relative to the sidebands caused, for example, by oscillations in the stabilization loop. Both processes adversely affect the performance of precision spectroscopy: the spectral resolution degrades and the excitation rate of the clock transition drops. The quantitative description of phase noise relies on the phase-noise power spectral density $S_\varphi(f)$ and on the related power spectral density of frequency fluctuations $S_\nu(f) = S_\varphi(f) \times f^2$. Obviously, both functions for the harmonic wave grow in proportion to $K^2$ [b].

[b] Although the analysis presented below is carried out for a continuous-wave laser, it applies equally to an individual mode of a femtosecond comb stabilized to the same ultrastable laser [40]. When the comb is phase-locked to the reference laser, the noise of each mode within the loop bandwidth reproduces its noise with the addition of the residual locking noise. Upon generation of the Kth harmonic, all combinations of modes contributing to a fixed spectral component of the harmonic add up with a summed phase, so that the phase-noise spectral density of the harmonic mode increases by a factor of $K^2$ — just as for continuous-wave radiation.

The spectrum of radiation with a fluctuating phase is related to the power spectral density of its frequency noise $S_\nu(f)$ by the relation [25–27]:

$$S_E(\nu-\nu_0) = E_0^2 \int_{-\infty}^{\infty} \exp(-i\cdot 2\pi(\nu-\nu_0)\tau) \times \\ \times \exp\left\{-2\int_0^{\infty} S_\nu(f)\cdot\frac{\sin^2(\pi f\tau)}{f^2}df\right\}d\tau. \qquad (16)$$

where $E_0^2$ is the total power of the light wave $E(t) = E_0\, exp(2\pi i\nu_0 t + i\phi(t))$, and $\nu_0$ is the central frequency of the radiation. For general forms of the frequency noise $S_\nu(f)$, no analytical form of the integral (16) is available. Two limiting cases for estimating the spectrum $S_E(\nu-\nu_0)$ [25–27] have been considered in the literature, based on the value of a parameter that characterizes the form of $S_\nu(f)$: $\Delta\nu_{\text{rms}} \times \tau_\nu$, where $\Delta\nu_{\text{rms}} = \langle\Delta\nu^2(t)\rangle^{1/2} = \left(\int_0^\infty S_\nu(f)df\right)^{1/2}$ is the root-mean-square spread of the instantaneous radiation frequency $\Delta\nu(t) = d\phi(t)/dt$, $1/\tau_\nu$ is the characteristic width of the distribution $S_\nu(f)$. When $\Delta\nu_{\text{rms}} \times \tau_\nu >> 1$, the instantaneous frequency acquires the physical meaning of a wandering radiation frequency, and the spectrum $S_E(\nu-\nu_0)$ takes a Gaussian form $S_E(\nu-\nu_0) \propto exp\left(-\frac{1}{2}\left(\frac{\nu-\nu_0}{\Delta\nu_{\text{rms}}}\right)^2\right)$ with a width of $\sim\Delta\nu_{\text{rms}}$. This is the typical case for stabilized lasers with flicker noise and technical frequency drifts. In the opposite limit, when $\Delta\nu_{\text{rms}} \times \tau_\nu << 1$, the spectrum $S_E(\nu-\nu_0)$ takes a Lorentzian form with a width of $(\Delta\nu_{\text{rms}})^2\tau_\nu$, which is noticeably smaller than $\Delta\nu_{\text{rms}}$. The authors of [25] extended the criterion $\Delta\nu_{\text{rms}} \times \tau_\nu <> 1$ to the components of the spectral density $S_\nu(f)$, each of which oscillates with a period of $1/f$. Frequency components for which the product of the amplitude and the oscillation period exceeds unity, $S_\nu(f) \times 1/f > 1$, are responsible for actual oscillations of the radiation frequency and directly form the spectral line. Conversely, frequency components for which this product is below unity, $S_\nu(f) \times 1/f < 1$, contribute little to the linewidth and form the wings of the line profile. This allowed the authors to formulate an estimate of the width of the spectrum $S_E(\nu-\nu_0)$:

$$\Delta\nu_{\text{FWHM}} = \sqrt{8\cdot\ln(2)\cdot A} \approx 2.3\times\sqrt{A}, \qquad (17)$$

where $A$ is the area under the curve of the frequency-noise spectral density $S_\nu(f)$ in the region extending up to the frequency of intersection of $S_\nu(f)$ with the so-called $\beta$-separation line: $S_\beta(f) = 8\cdot\ln(2)\cdot f/\pi^2 \approx 0.56f$. The lower integration limit is set by the spectrum observation time $T_o$: components of $S_\nu(f)$ with $f < 1/T_o$ do not contribute to the observed linewidth and manifest themselves as a drift of the central frequency.

As already noted, in the weak-conversion regime the resulting harmonic field is the Kth power of the input field. This relation also applies to the field of an individual mode in the spectrum of a femtosecond frequency comb. In the Kth harmonic, the phase and the instantaneous frequency increase by a factor of K, i.e., for the power-spectral-density components we have $S_\nu^{(K)}(f) = K^2 S_\nu(f)$. Under these conditions, the criterion separating the limiting cases of spectrum formation changes to $K\Delta\nu_{\text{rms}} \times \tau_\nu <> 1$, while the recipe for estimating the spectral width from the new $S_\nu^{(K)}(f)$ remains valid.

The application of the technique of Ref. [25] is illustrated in Figure 2 for the fundamental radiation and its eighth harmonic. The function $S_\nu(f)$ of the fundamental radiation was constructed from the data of Refs. [28,29] for an ultrastable laser with a reference cavity, based on experimental measurements and a fit by a set of analytical components.

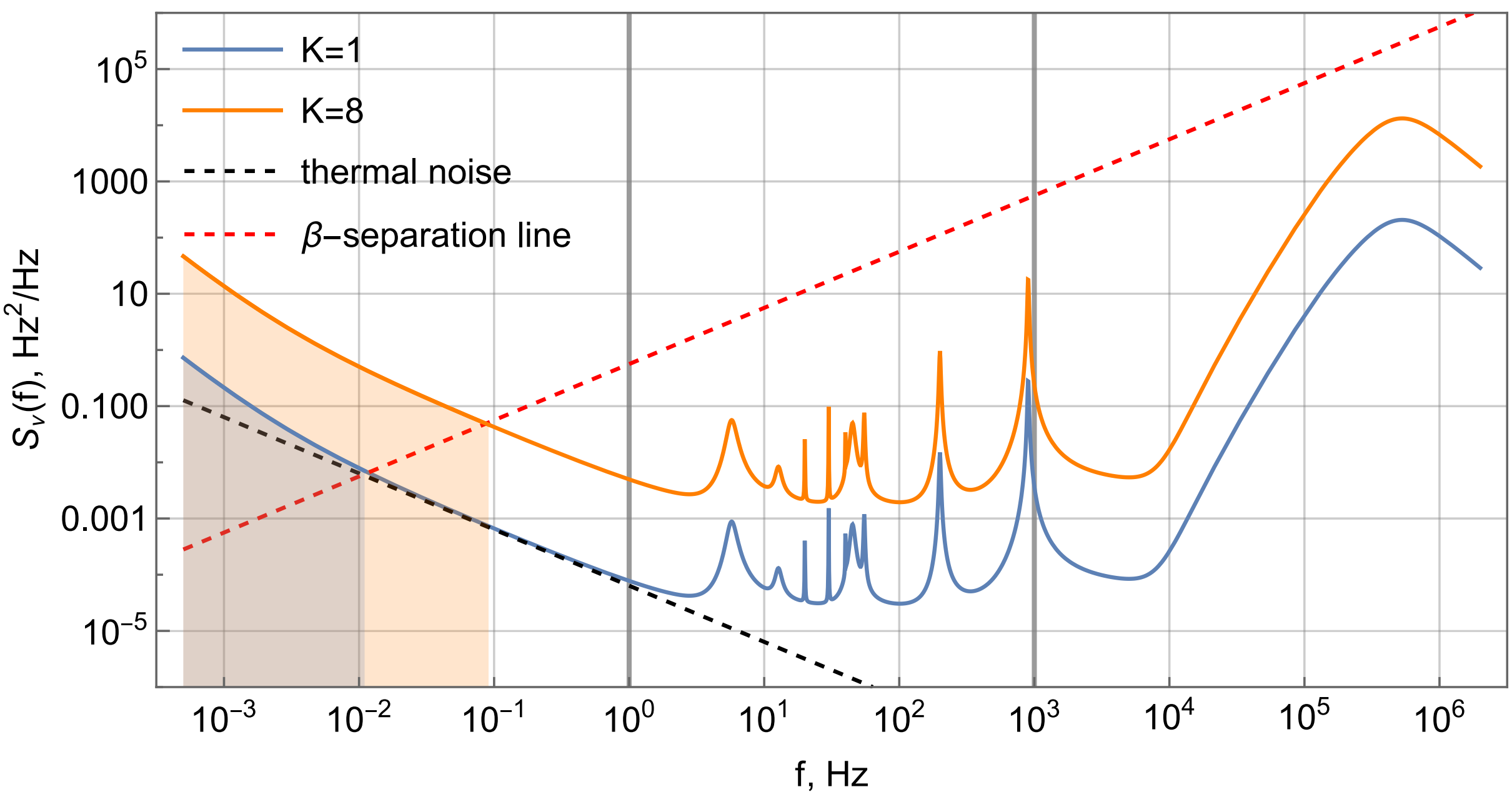

*Figure 1. Power spectral density of the frequency noise of an ultrastable laser (blue curve), constructed from the fitted data of [28,29] and the result of its rescaling to the eighth harmonic (K=8, multiplication by $K^2$=64). The red dashed line is the β-separation line; the black dashed line is the thermal-noise limit of the cavity. The shaded areas determine the spectral linewidths of the original laser and of the harmonic according to (17). The left boundary of the integration region is chosen depending on the averaging time.*

At low frequencies ($f < 1\ Hz$), the noise of such lasers is dominated by flicker frequency noise $\propto h_{-1}/f$, associated with the thermal noise of the reference cavity, and by random-walk frequency noise $\propto h_{-2}/f^2$ ($h_{-1}$ and $h_{-2}$ are fit coefficients). The thermal noise of the reference cavity originates from thermal fluctuations of the mirror surfaces, which lead to instability of its length. This noise is fundamental in nature and sets the ultimate frequency instability of a laser stabilized to an external cavity. In the intermediate frequency range ($1\,Hz < f < 10^3\ Hz$), the noise spectrum is described by a pedestal corresponding to white frequency noise $\propto h_0 f^0$, on which narrow peaks caused by vibrations of seismic and acoustic origin are superimposed. At high frequencies ($f > 10^3\ Hz$), the dominant noise contribution is a characteristic peak at the unity-gain frequency of the laser frequency-stabilization servo loop (servo bumps). In the laser emission spectrum, servo bumps appear as broad sidebands symmetrically offset from the carrier. The intersection of the $\beta$-separation line with the curve of $S_\nu(f)$ occurs in the flicker-noise region. The curve of $S_\nu^{(K)}(f)$ is plotted in accordance with the transformation $S_\nu^{(K)}(f) = K^2 S_\nu(f)$, i.e., on a logarithmic scale it is “raised” by $\log(K^2)$. The intersection of this curve with the $\beta$-separation line still occurs in the flicker-noise region. The growth of the area under the curve of $S_\nu^{(K)}(f)$ up to the intersection with the $\beta$-separation line is determined mainly by the rise of the amplitudes, i.e., the spectral linewidth of the harmonic grows in proportion to the harmonic order. This result is consistent with the picture of the transformation of frequency wandering upon harmonic generation: the range of frequency excursions increases by a factor of K, and the spectral shape remains Gaussian. One may expect that if the intersection of $S_\nu^{(K)}(f)$ with the $\beta$-separation line occurred in a region dominated by white frequency noise, the contribution of the Lorentzian wings to the spectrum would increase and the linewidth scaling would tend to change from K to $K^2$.

We note that in Ref. [20] a quadratic growth of the beat-note linewidth with harmonic order was observed experimentally. In that case, the beat-note width was determined not by the noise of the master laser (common to both arms) but by the technical Doppler noise of the interferometer path-length difference [30], which produced a Lorentzian profile and quadratic growth; after active stabilization of the path-length difference, the broadening disappeared.

An important feature of the spectrum of ultrastable lasers is the servo bumps - regions of frequency detuning where the phase lag of the feedback signal reaches $\pi$ and the loop amplifies the noise instead of suppressing it [31]. This transfers part of the radiation power from the carrier into two symmetric broad peaks in the vicinity of frequencies set by the feedback-loop bandwidth. The prominence of the servo bumps depends on the type of laser and stabilization loop. It is higher for lasers with a broad intrinsic linewidth, which require high gain and a wide loop bandwidth - primarily for external-cavity diode lasers (ECDLs), considered below. In contrast, for fiber lasers and long-cavity ECDLs, which have substantially lower intrinsic frequency noise, the required loop bandwidth and gain are much lower, and servo bumps may not appear in the spectrum at any appreciable level. The estimates given below should be regarded as characteristic of the former case.

The fraction of power contained in the carrier relative to the total radiation power of a continuous-wave laser can be estimated from the formula:

$$\frac{P_{carr}}{P_{total}} = e^{-\varphi_{rms}^2}, \tag{18}$$

where $\varphi_{rms}$ is the root-mean-square phase excursion, given by the expression:

$$\varphi_{rms}^2 = \int_{1/T_o}^{+\infty} S_\varphi(f)df. \tag{19}$$

For the Kth harmonic, $\varphi_{rms} \to K\varphi_{rms}$, and

$$\frac{P_{carr}}{P_{total}} = e^{-K^2\varphi_{rms}^2}. \tag{20}$$

To study the influence of the servo bumps, it suffices to restrict the measurement time $T_o$ to no more than 1 ms, since servo bumps usually appear at frequencies well above 1 kHz. Typical values of $\varphi_{rms}$ for laser frequency-stabilization systems are of the order of 10-100 mrad [32], which corresponds to more than 99% of the radiation power being contained in the carrier. In higher harmonics this balance is inevitably upset, leading to depletion of the carrier, up to its collapse when the argument of the exponent in (20) reaches the threshold value $(1\ rad)^2$ [33,34].

To assess the influence of the servo bumps on the carrier power fraction in the process of nonlinear harmonic generation, we measured the phase-noise power spectral density of an ECDL stabilized to a vacuum-housed Fabry–Pérot reference cavity [32]. A simplified measurement scheme is shown in Figure 3. A radio-frequency signal whose phase noise corresponds to the phase noise of the laser is formed at a photodetector as a heterodyne beat between the diode-laser radiation at the stabilized laser output and the radiation of the same laser transmitted through the cavity. The width of the transmission peak of the Fabry–Pérot cavity is 8.3 kHz, which makes it possible to suppress the high-frequency noise (including the servo bumps). The transmitted radiation is injected into a laser diode to increase power. The amplified laser beam is intended for exciting the clock transition in the ytterbium-171 ion [35]. The frequency detuning between the filtered and unfiltered radiation was provided by additional acousto-optic modulators (AOMs). The phase noise was measured with a Rohde & Schwarz FSW analyzer. To demonstrate the influence of the feedback-loop settings, measurements of $S_{\varphi}(f)$ were performed for different values of the gain setting. The results are shown in Figure 4. Figure 5 shows the calculated dependence of $\varphi_{rms}$ on the gain. Finally, for each value of the gain setting, the dependence of the carrier power fraction on the harmonic order was calculated (Figure 6).

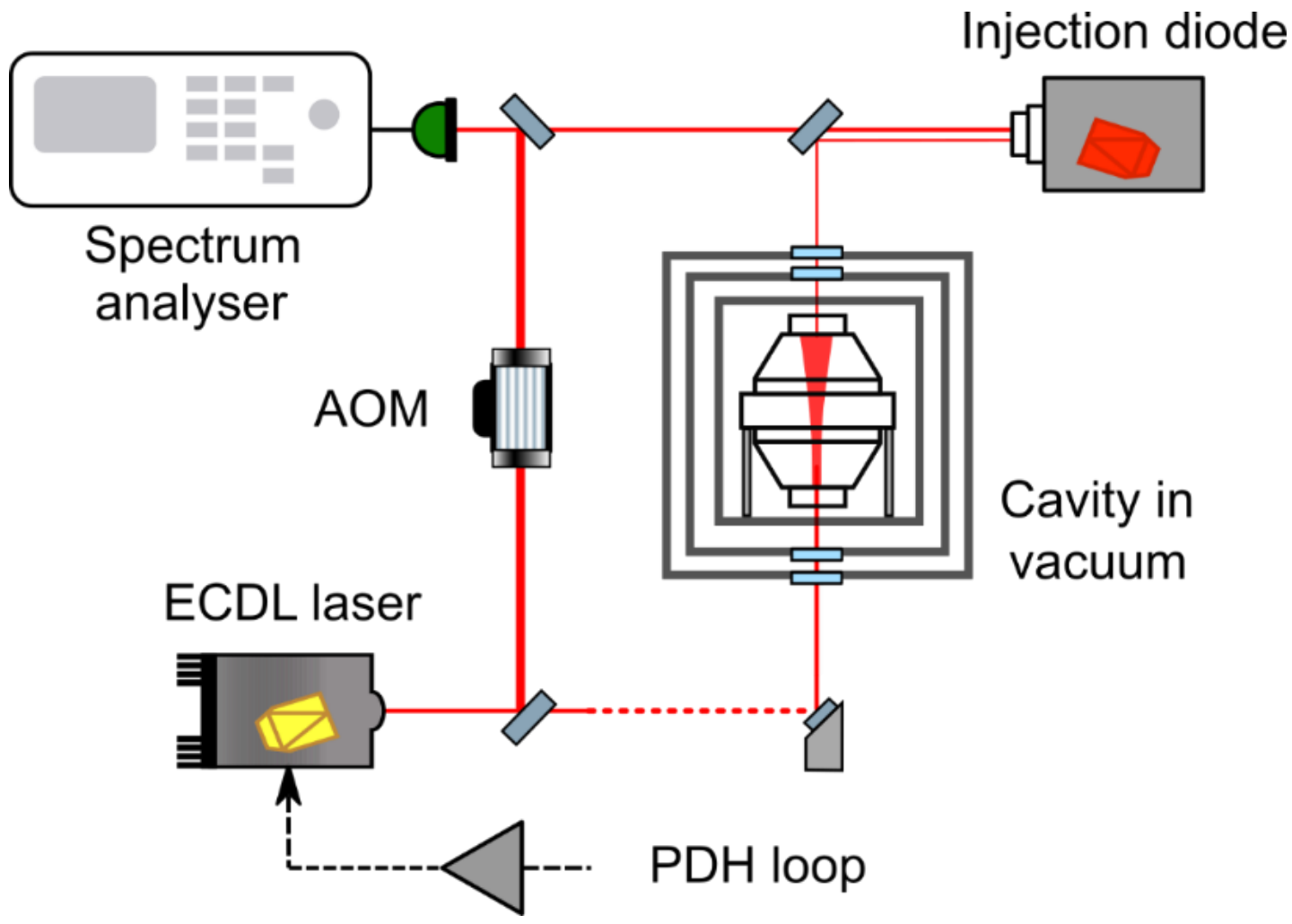

*Figure 3. Simplified optical layout for measuring the phase noise of a laser stabilized to a reference cavity. The beat note between the stabilized radiation at the laser output and the radiation transmitted through the cavity is recorded with a spectrum analyzer. In the transmitted radiation, the servo bumps are filtered out by the cavity transmission bandwidth. The beat-note frequency corresponds to the offset produced by the AOM.*

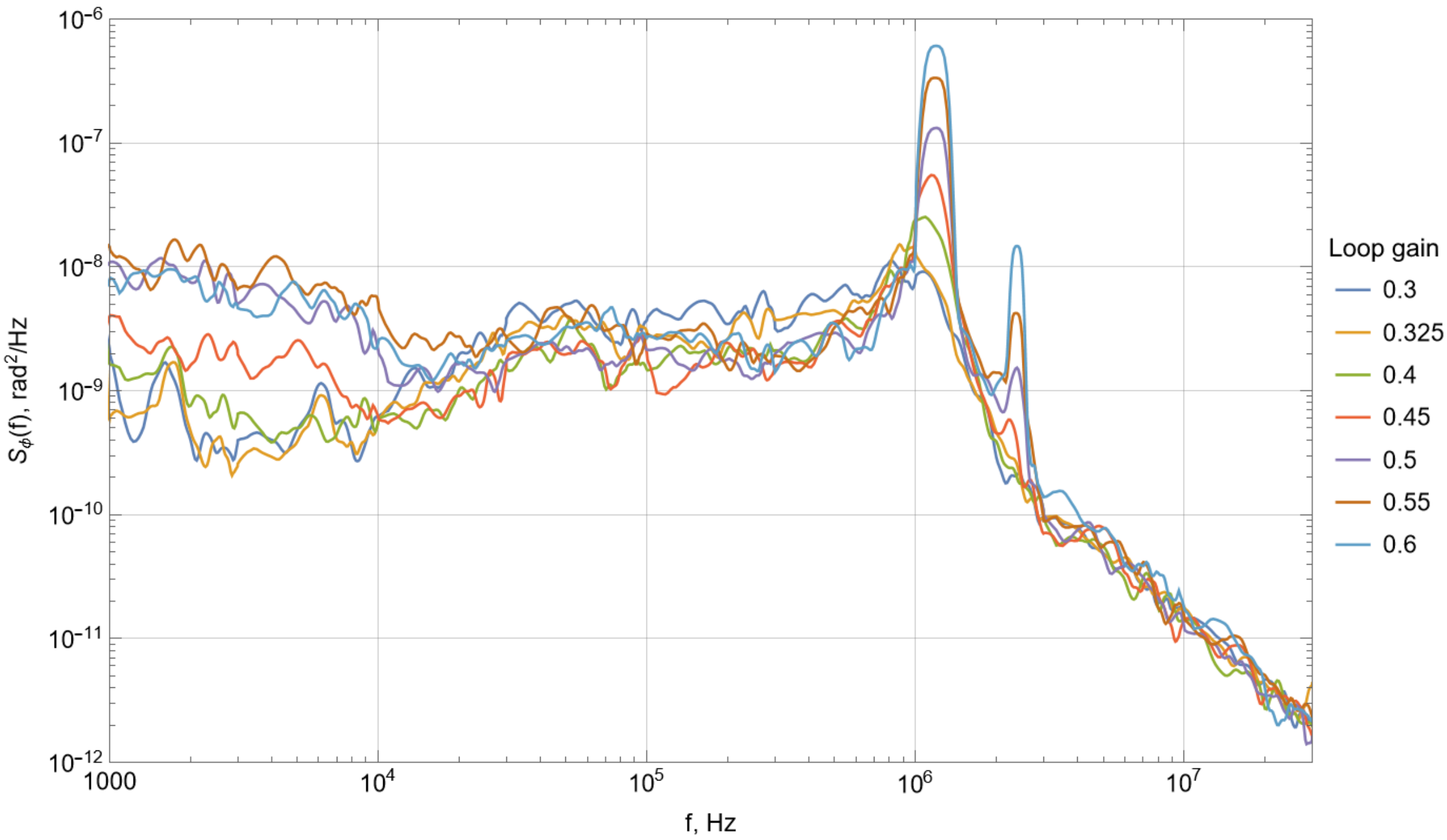

*Figure 4. Phase-noise power spectral density curves measured for different values of the servo-controller gain setting in the PDH loop. When the gain is doubled, the servo-bump peak grows by approximately two orders of magnitude.*

Increasing the servo-controller gain raises the phase noise at the servo-bump frequency by almost two orders of magnitude. The root-mean-square phase excursion then grows fourfold. Even at the optimal loop gain, the carrier power fraction at the eighth harmonic is less than one half. For harmonics above the 15th, less than 10% of the power remains in the carrier. When the gain is doubled, carrier collapse occurs already at the 2nd-3rd harmonic.

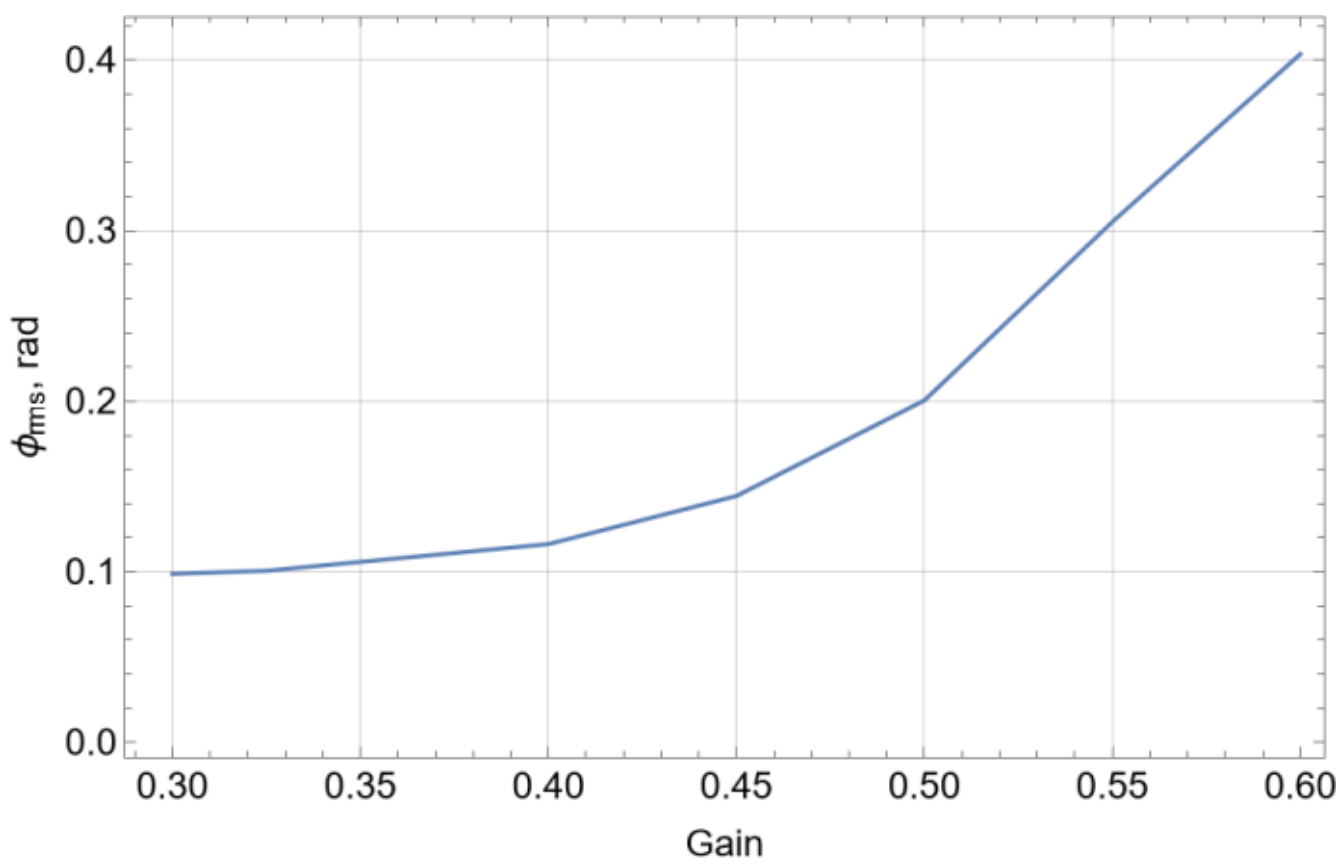

*Figure 5. Root-mean-square phase excursion as a function of the servo-controller gain setting.*

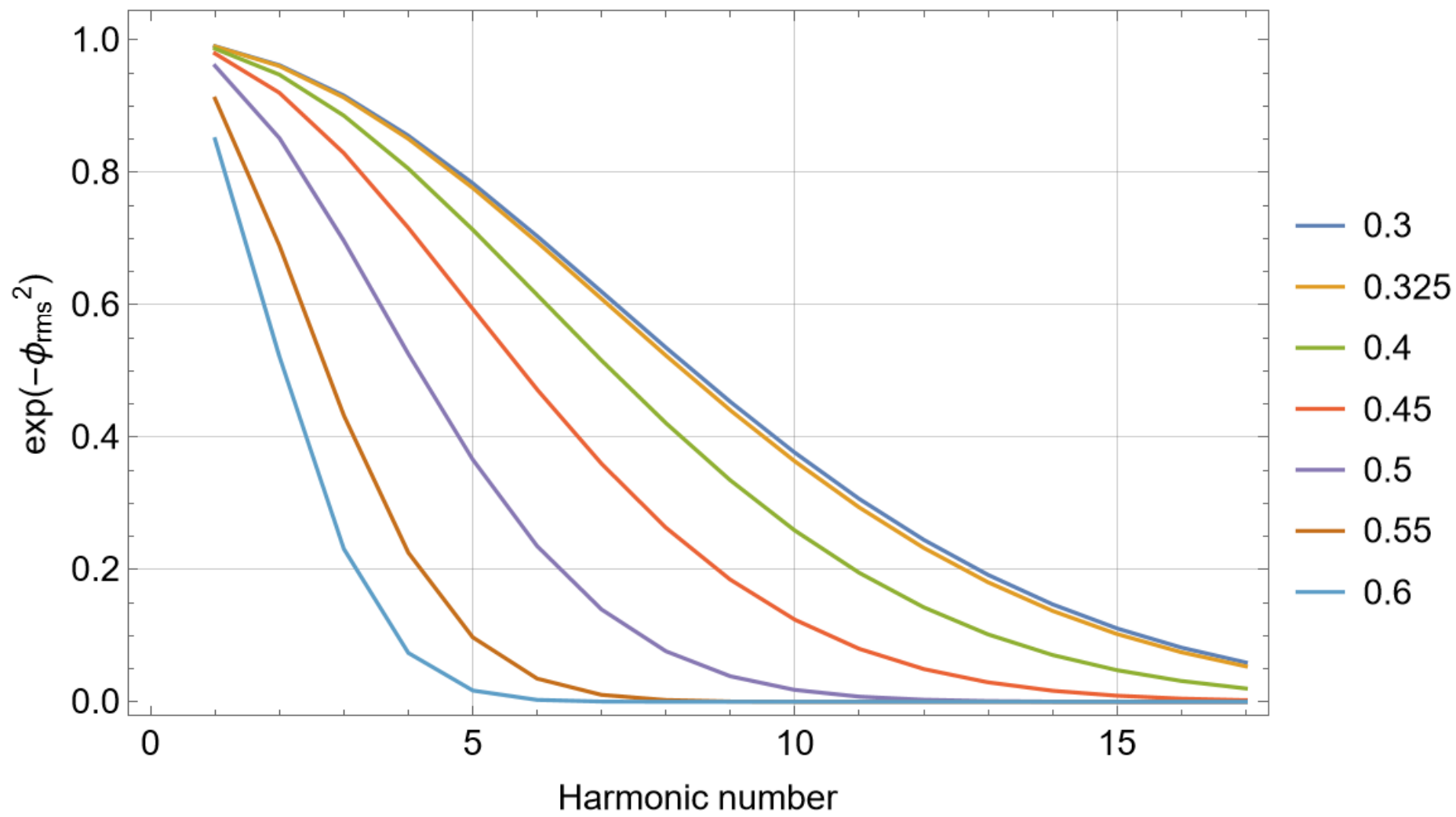

*Figure 6. Carrier power fraction as a function of harmonic order for different values of the gain setting.*

A laser with a linewidth as narrow as a few hertz may lose a significant fraction of its coherent power upon generation of higher harmonics. For excitation of the nuclear transition in thorium-229, which requires generation of the 5th-8th harmonic of infrared radiation, control of the phase noise at the unity-gain frequency of the servo loop becomes no less important than achieving a narrow linewidth. Possible ways of suppressing the servo bumps are filtering of the radiation with a high-finesse cavity [36,37] followed by amplification via injection locking [35], as well as direct phase-noise compensation (feedforward) techniques [38,39].

As applied to the comb, in expression (19) $S_\varphi(f)$ should be understood as the sum of the transferred noise of the reference laser and the residual noise of the comb phase-locking loop, including its own servo bumps. Owing to the exponential dependence (18), the power fraction in the carrier mode of the harmonic is determined by the largest of the root-mean-square phase excursions; thus, the reference-laser characteristics measured in this work set a lower bound on the carrier degradation for both types of sources considered in Section II.

## IV. Conclusion

We considered two aspects of frequency multiplication of stabilized lasers that determine the suitability of the resulting radiation for precision spectroscopy: the power in the useful spectral component and the preservation of spectral purity.

Regarding the first aspect, it has been shown that in the undepleted-pump approximation the power of an individual mode of the Kth harmonic of a femtosecond comb at K=2 coincides with the harmonic power of a continuous-wave laser of equal average power, and at $K \geq 3$ exceeds it, growing by a factor of approximately N (N is the number of comb modes, $\sim 10^5$ for typical parameters) with each next harmonic order: the distribution of power over the modes is compensated by the gain in peak intensity. First-order dispersion of the medium narrows the spectral conversion bandwidth but does not affect the central mode of the harmonic, whose power continues to grow quadratically with the medium length.

Regarding the second aspect, a substantial asymmetry between the two effects of phase-noise multiplication has been established. The spectral linewidth grows linearly with the harmonic order for lasers whose spectrum is determined by the flicker frequency noise of the reference cavity, which is characteristic of modern ultrastable systems. In contrast, the power fraction in the carrier decreases exponentially (20) and is determined by the high-frequency noise in the region of the unity-gain frequency of the stabilization loop. Measurements of the phase-noise spectral density of a laser stabilized to a Fabry–Pérot cavity have shown that at a root-mean-square phase excursion of ~100 mrad, typical of an optimally tuned loop, carrier collapse occurs in the vicinity of the 10th harmonic. Since the $K^2$-fold multiplication of phase noise is a property of the frequency-conversion process itself, independent of its efficiency and of the type of primary source, the estimates obtained apply to existing VUV sources for spectroscopy of the nuclear transition in thorium-229 (148.4 nm), which use generation of the seventh [18], eighth [16], and sixteenth [17] harmonics.